\documentstyle[epsfig,longtable]{aipproc}

\begin{document}
\title{Longitudinal Momentum Fraction ${X_{L}}$\\ for Two High $P_{t}$ Protons in
pp$\rightarrow$ppX Reaction}

\author{  D. Zhalov$^e$, 
  \underline{  S. Heppelmann$^e$}, 
  J. Alster$^a$, G. Asryan$^{c,b}$,
\newline 
  Y. Averichev$^{h}$, D. Barton$^c$, V. Baturin$^{e,d}$, N. Bukhtoyarova$^{c,d}$, 
\newline
  A. Carroll$^c$,
  T. Kawabata$^f$, 
  A. Leksanov$^e$, 
  Y. Makdisi$^c$, 
  A. Malki$^a$,	
\newline
E. Minina$^e$, I. Navon$^a$, 
  H. Nicholson$^g$, A. Ogawa$^e$, 
  Yu. Panebratsev$^{h}$, 
  E. Piasetzky$^a$,
  A. Schetkovsky$^{e,d}$, 
   S. Shimanskiy$^{h}$, 
  A. Tang$^i$,
\newline
   J.W. Watson$^i$,
   H. Yoshida$^f$
}

\address{$^a$School of Physics and Astronomy, Sackler Faculty of Exact Sciences, Tel Aviv University, Ramat Aviv 69978, Israel \\
 $^b$Yerevan Physics Institute, Yerevan 375036, Armenia \\ 
 $^c$Collider-Accelerator Department, Brookhaven National Laboratory,Upton, NY 11973, USA \\
 $^d$Petersburg Nuclear Physics Institute, Gatchina, St. Petersburg 188350, Russia \\
 $^e$Physics Department, Pennsylvania State University,University Park, PA 16801, USA 
 $^f$Department of Physics, Kyoto University, Sakyoku, Kyoto, 606-8502, Japan \\
 $^g$Department of Physics, Mount Holyoke College, South Hadley, MA 01075, USA \\
 $^h$J.I.N.R., Dubna, 141980, Russia \\ 
 $^i$Department of Physics, Kent State University, Kent, OH 44242, USA}
\maketitle

\begin{abstract}
We present an analysis of new data from Experiment E850 at BNL. We have characterized the inclusive cross
section near the endpoint for pp  exclusive scattering in Hydrogen and in Carbon
with incident beam energy of 6 GeV. We select events with a pair of back-to-back
hadrons at large transverse momentum. These cross sections are parameterized with a form $\frac{d \sigma}{ d X_{L}}$
$\sim(1-X_{L})^{p}$, where ${X_{L}}$ is the ratio of the longitudinal momentum of the observed pair to the total incident
beam momentum. Small value of $p$ may suggest that the number of partons participating in the reaction is large and reaction has a strong dependence on the center-of-mass energy. We also discuss nuclear effects observed in our kinematic region.
\end{abstract}
%
%
%
%
\par
There are two main questions which can be addressed  in the study  of the Longitudinal
Momentum Fraction carried by a pair of large $P_{t}$ final state particles in the pp$\rightarrow$ppX reactions.
We learn first something about the number of partons involved in the interaction.  Concluding, as we will,
that the number of partons involved is large, we discuss a mechanism for picking up a momentum boost from
the spectator nucleus in events where the observed momentum fraction is equal and larger than the incoming momentum.

We defined the Longitudinal Momentum Fraction variable ${X_{L}}$ as:
\begin{equation}
\small
	{X_{L}}=\frac{P_{L_{1}}+P_{L_{2}}}{P_{L_{inc}}}.
\end{equation}
Here $P_{L_{inc}}$ is the momentum of the beam coming along the Z-axis,
$P_{L_{1}}$ and $P_{L_{2}}$ are the longitudinal components of the two detected out-going 
particles having large transverse momenta.
We have argued that the two detected out-going particles in the pp$\rightarrow$ppX reaction are protons \cite{malkiprl}.
We then expect the shape of the cross section, as $X_{L}$ approaches unity, to reflect the number of  partons
participating in the collision.  If the cross section is parameterized as 
\begin{equation}
\frac{d\sigma}{d{X_{L}}} \sim (1-X_{L})^p
\end{equation}
then a large $p$ is expected for  single parton collisions, reflecting the endpoint behavior of structure
functions and
fragmentation functions \cite{brodsk95}. As the number of participating partons increases, we expect the power $p$ to be 
reduced.

In the experiment discussed here, we measure $p$ for selected classes of two track events. Our result, 
that $p$ is in the range of 1-2, indicates that many partons are participating in the hard process.  
It can be expected that the cross section may be
falling very rapidly with  center-of-mass (cm) energy for these events, perhaps similar to the pp elastic energy dependence.

The cm energy dependence of the underlying pp$\rightarrow$ppX cross section is affected by the
number of partons participating in the interaction.  The underlying cross section falls with energy 
as a power of  energy, and that power increases
as the number of partons increases.  It is known that for exclusive scattering, the cross section has a dependence on the cm energy given by (s is a Mandelstam variable)
$\sigma \sim s^{-m} ,       m=(n_{1}+n_{2}+n_{3}+n_{4}-2) $
for a process with $n_{1}+n_{2}$ partons $\rightarrow$ $n_{3} + n_{4}$ partons \cite{farrar73}.

With this observation about the nature of the selected  pp$\rightarrow$ppX events, we can further conclude that interesting
nuclear effects would also be expected. The nucleus can change the cm energy of the incident proton and
the nuclear proton by virtue of nuclear fermi momentum.  If the nucleus contributes a target proton 
which is moving in the direction of the beam particle, the cm energy is reduced.  Because the increase 
in cross section associated with a reduction in cm energy is very large, we can expect a strong bias 
for the nucleus to contribute momentum to the scattering process in the forward direction. The 
typical magnitude of this nuclear contribution to the final state momentum may be much larger than 
typical fermi momentum seen in other processes.

Since the nucleus tends to add positive momentum to the two final state particles, in some sense, 
the stopping power for these events is apparently negative.  When one compares the same pp$\rightarrow$ppX 
processes measured in nuclear matter with that measured in Hydrogen, the effect of the nucleus is to
move strength in the $X_{L}$ distribution to larger $X_{L}$.  If a ratio, like a transparency
ratio of the cross section in nuclear matter to that measured in Hydrogen is calculated, 
the ratio will be artificially enhanced.

The data  used for this analysis was collected on two targets $^{12}C$ and $CH_{2}$ using the EVA 
Spectrometer (E850 Exp., BNL, AGS ). The $CH_{2}$ target 
enabled us to extract cross sections for free proton (Hydrogen) scattering.
The acceptance of the detector for 
two high $P_{t}$ tracks in the r-z plane was centered around the opening angle for 
elastic pp$\rightarrow$pp scattering at $90^{\circ}$ in the cm frame \cite{leksanov00}.
Charged particle acceptance in the azimuthal angle was nearly 2$\pi$. 

\begin{figure}[h!]
\centering
\mbox{{\epsfig{figure=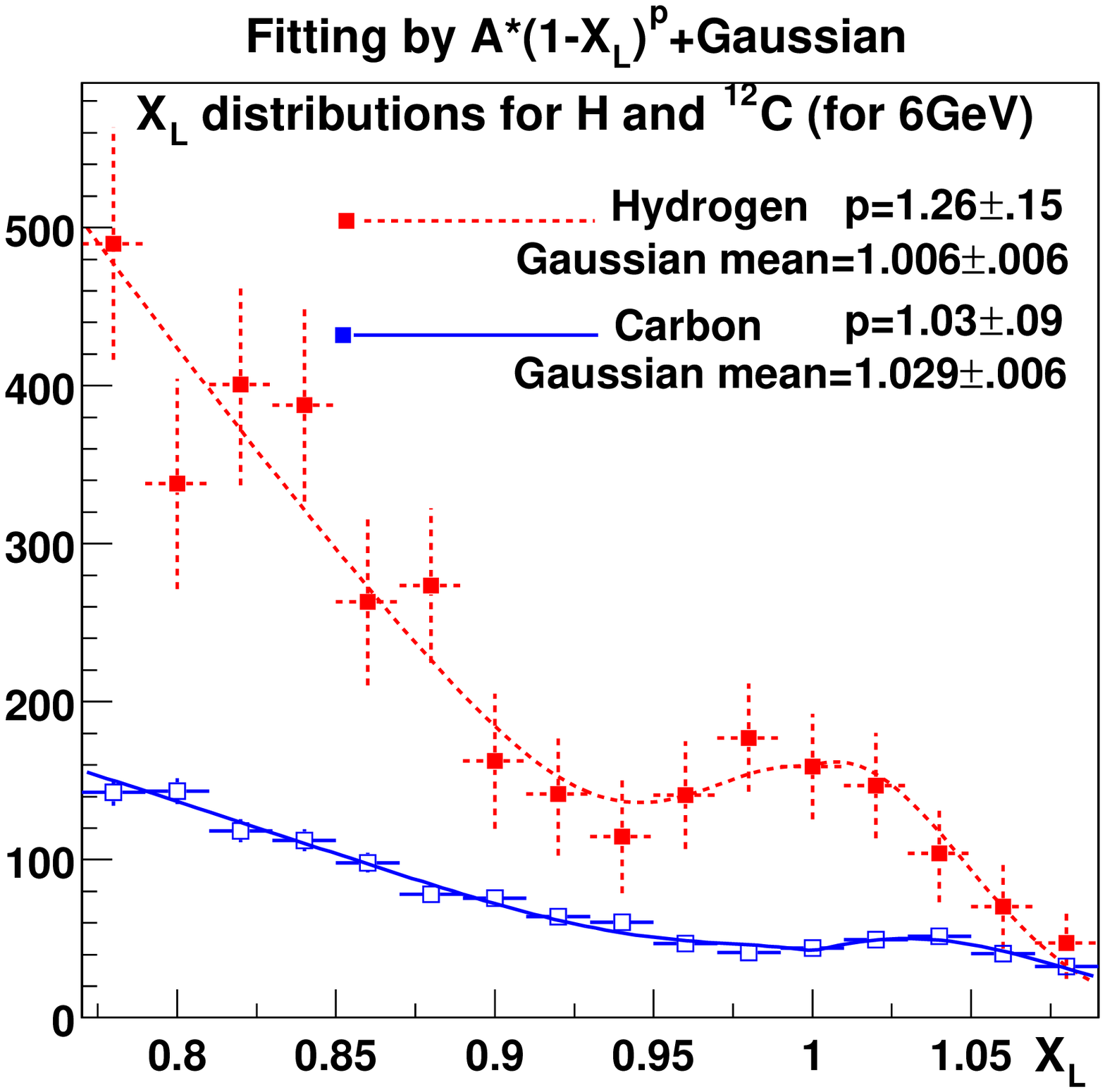, height=7.5cm,width=7.5cm}}\quad
	{\epsfig{figure=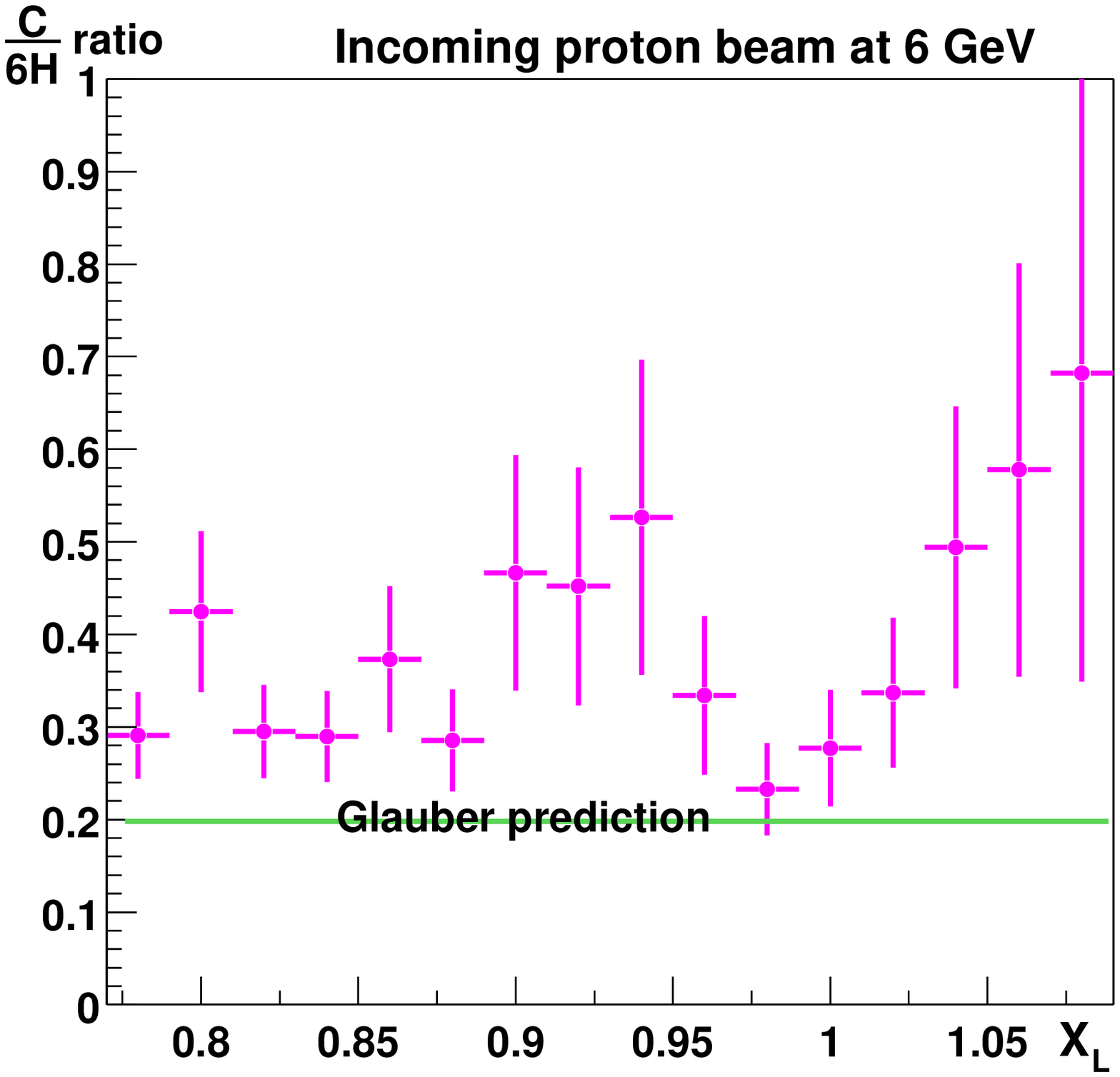, height=7.5cm, width=7.5cm}}}
\caption{LEFT: ${X_{L}}$ distributions for Carbon nucleus and 6*(Hydrogen) with two and only two tracks cut. RIGHT: Ratio of the Carbon signal to the 6*(Hydrogen) signal with two and only two tracks cut.}
\label{fig1}
\end{figure}

In Fig. \ref{fig1}, we present the ${X_{L}}$ distribution for Carbon and Hydrogen. In one 
case we applied a 'semi-elastic' cut (two and only two charged tracks in our charged track
acceptance) which allowed us to look at the ${X_{L}}$ distribution for 
pp$\rightarrow$pp events with a background of pp$\rightarrow$pp+$X^{0}$ and
pp$\rightarrow$pp+$X^{charged}$. The quasi-elastic signals have been extensively
analyzed elsewhere \cite{leksanov00,carroll88}.

Secondly, we select a more inclusive data set made up of
events with two high ${P_{t}}$ tracks but any number of additional
lower momentum charged tracks in the pp$\rightarrow$ppX process.  
\par

We have fitted  ${X_{L}}$ distribution with a form (here A and p are the fit constants):
\begin{equation}
	\frac{d\sigma}{d{X_{L}}}=A*(1-{X_{L}})^{p}+Gaussian
\end{equation}
The polynomial part of the fit was restricted for ${X_{L}}<$1 while the 
Gaussian is included
to account for the quasi-elastic contributions near ${X_{L}}\simeq 1$.

The peak corresponding to the Gaussian part of the fit is actually shifted 
for Carbon events from  ${X_{L}}=1$ towards larger ${X_{L}}$ because of the $s^{-10}$ 
dependence of the pp$\rightarrow$pp reaction's cross section.

We have determined the power of the polynomial in the fit to be in the range 1.1-1.8 
depending on the tightness of the cuts around the region of the elastic kinematics.
According to our interpretation, we conclude that the 
number of quarks participating in the process is large.
Consequently, the power {m} of $s^{-m}$ dependence of the cross section would be expected
to be large as well.

In Fig. \ref{fig1}, the effective ratio for scattering in Carbon vs Hydrogen is shown.
While the ratio is well in excess of Glauber levels, our interpretation is that the Carbon
distribution is enhanced due to the strong energy dependence of the cross section.

A recent study of neutron correlations from this experiment (E850 collaboration at BNL) has 
shown an anomalously large yield of neutrons going into the backward hemisphere in 
pp$\rightarrow$ppX reactions compared to the yields reported by other experiments \cite{malkiprl,malki00}. 
This result was observed with a data sample similar to the one described above.
A possible interpretation of this phenomenon is that strong s-dependence of the cross-section for this
class of events may select events with large positive fermi momentum. The nucleus must balance
this momentum by some kind of recoil in the backward direction.
The idea that most of the recoil momentum is absorbed by a single recoiling neutron leads
to the prediction of enhanced backward neutron production \cite{frankfurt77}. 
Since a large yield of backward neutrons has been 
observed in the  pp$\rightarrow$ppX data set described here, the picture seems to have merit.
\par
Keeping this in mind, we looked at the ${X_{L}}$ distribution for Carbon and Hydrogen (Fig. \ref{fig1}). 
In the case of exclusive events with two high $P_{t}$ particles, we can clearly see the shift in the position 
of the Gaussian peak in the Carbon sample. This effect 
is known to be caused by the ${s}^{-10}$ dependence of the cross section. 
We have compared the more inclusive data set distributions for Carbon and Hydrogen 
for pp$\rightarrow$ppX and interpreted
the enhanced ratio as evidence for a similar shift in the Carbon distribution with respect to the Hydrogen 
for this more inclusive sample as well.




\begin{references}

\bibitem{malkiprl}Malki A. et al, {\it to be published}.
\bibitem{brodsk95}Brodsky S. et al, {\it Nucl. Phys.}\ {\bf B441}, 197 (1995).
\bibitem{farrar73}Brodsky S., Farrar G., {\it Phys. Rev.\ Lett.}\ {\bf 31}, 1153 (1973).
\bibitem{leksanov00}Leksanov A. et al, in {\it Proceedings of CIPANP2000, Quebec, May 22-28, 2000}.
\bibitem{carroll88}Carroll A. et al, {\it Phys. Rev.\ Lett.}\ {\bf 61}, 1698 (1988).
\bibitem{malki00}Malki A. et al, in {\it Proceedings of CIPANP2000, Quebec, May 22-28, 2000}.
\bibitem{frankfurt77}Frankfurt L.L., Strikman M.I., {\it Phys. Lett.}\ {\bf B69}, 87 (1977).
\end{references}
\end{document}